\documentclass[useAMS]{mn2e}
\usepackage{graphicx}

\def\simgt{\ {\raise-.5ex\hbox{$\buildrel>\over\sim$}}\ }
\def\I{\'\i}
\def\cd{cd$^{-1}$\,}
\def\kms{kms$^{-1}$\,}

\begin{document}

\title[Simultaneous p and g mode pulsation of HD 8801 and $\gamma$ Peg]
{Confirmation of simultaneous p and g mode excitation in HD 8801 and
$\gamma$ Peg from time-resolved multicolour photometry of six candidate 
"hybrid" pulsators}
\author[G. Handler]
 {G. Handler
\and \\
Institut f\"ur Astronomie, Universit\"at Wien,
T\"urkenschanzstrasse 17, A-1180 Wien, Austria}

\date{Accepted 2008 July 17.
 Received 2008 August 13;
in original form 2008 September 10} 
\maketitle 
\begin{abstract} 
We carried out a multi-colour time-series photometric study of six stars
claimed as "hybrid" p and g mode pulsators in the literature. $\gamma$~Peg
was confirmed to show short-period oscillations of the $\beta$~Cep type
and simultaneous long-period pulsations typical of Slowly Pulsating B
(SPB) stars. From the measured amplitude ratios in the Str\"omgren uvy
passbands, the stronger of the two short period pulsation modes was
identified as radial; the second is $\ell=1$. Three of the four SPB-type
modes are most likely $\ell=1$ or 2. Comparison with theoretical model
calculations suggests that $\gamma$~Peg is either a $\sim
8.5$\,M$_{\odot}$ radial fundamental mode pulsator or a $\sim
9.6$\,M$_{\odot}$ first radial overtone pulsator. HD~8801 was corroborated
as a "hybrid" $\delta$~Sct/$\gamma$~Dor star; four pulsation modes of the
$\gamma$~Dor type were detected, and two modes of the $\delta$~Sct type
were confirmed. Two pulsational signals between the frequency domains of
these two known classes of variables were confirmed and another was newly
detected. These are either previously unknown types of pulsation, or do
not originate from HD~8801. The O-type star HD~13745 showed
small-amplitude slow variability on a time scale of 3.2~days. This object
may be related to the suspected new type of supergiant SPB stars, but a
rotational origin of its light variations cannot be ruled out at this
point. 53~Psc is an SPB star for which two pulsation frequencies were
determined and identified with low spherical degree. Small-amplitude
variability was formally detected for 53~Ari but is suspected not to be
intrinsic. The behaviour of $\iota$~Her is consistent with non-variability
during our observations, and we could not confirm light variations of the
comparison star 34~Psc previously suspected. The use of signal-to-noise
criteria in the analysis of data sets with strong aliasing is critically
discussed.
\end{abstract}

\begin{keywords}
stars: variables: other -- stars: variables: $\delta$ Scuti -- stars: 
oscillations -- stars: individual: $\gamma$~Peg, 53~Psc, HD~8801, 
HD~13745, 53~Ari, $\iota$~Her, 34~Psc -- stars: early-type -- stars: -- 
chemically peculiar
\end{keywords}

\section{Introduction}

Stars can self-excite observable pulsations if a driving mechanism
operates in a resonant cavity in their interior or on their surface. This
condition is fulfilled in certain parts of the HR diagram, causing the
presence of different instability strips. A schematic representation of
the domains of some confirmed classes of pulsating variable stars is given
in Fig.\,1.

\begin{figure}
\includegraphics[width=88mm,viewport=05 05 360 345]{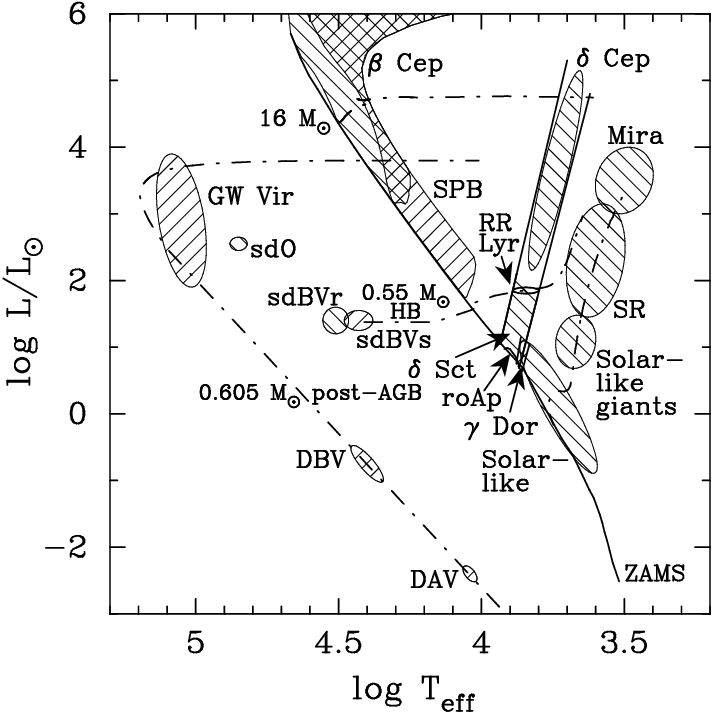}
\caption[]{Theoretical HR diagram schematically showing the locations of
different confirmed types of pulsating stars. Areas hatched from lower
left to upper right depict domains of g-mode pulsators, areas hatched from
lower right to upper left delineate domains of g-mode pulsators;
overlapping areas may contain "hybrid" pulsators. Parts of model
evolutionary tracks for main sequence, horizontal branch and post-AGB
stars are shown as dashed-dotted lines for orientation. Adapted and
updated from Christensen-Dalsgaard (2004).}
\end{figure}

The different types of pulsators have historically been classified
phenomenologically, which has usually later turned out to have a
physical reason. The individual classes can be separated in terms of
types of excited pulsation modes, mass and evolutionary state. As can be
seen in Fig.\ 1, the instability domains of some types of pulsators are
not fully distinct. Stars having two different sets of pulsational
mode spectra excited simultaneously may therefore exist within overlapping
instability strips. This is good news for asteroseismology -- the study of
stellar interiors based on modelling their pulsational spectra -- as the
information carried by both types of oscillations can be exploited.

In the past, there have been some reports of stars belonging to two
different types of pulsating variables (Chapellier et al.\ 1987, Mathias
\& Waelkens 1995), but the first systematic survey for such "hybrid"
pulsators was carried out by Handler \& Shobbrook (2002). This search was
successful in the discovery of both $\gamma$~Dor-type g modes and
$\delta$~Sct-type p modes in the star HD 209295 (CK~Ind).

However, Handler et al.\ (2002) showed that at least some of the g modes
of CK~Ind may have been excited through tidal effects from a close
companion in an eccentric orbit. Following this work, Henry \& Fekel
(2005) discovered both $\gamma$~Dor and $\delta$~Sct-type pulsations
in the single Am star HD 8801, and the MOST satellite found two additional
"hybrid" pulsators (HD~114839, King et al.\ 2006; BD+18~4914, Rowe et al.\
2006) both of which are again Am stars. This is in itself interesting as
Am stars have a lower incidence of pulsational variability than chemically
"normal" A/F main sequence stars (e.g., see Kurtz 1989).

Among the B-type stars, "hybrid" SPB/$\beta$~Cep pulsations have been
reported for several objects (e.g., see Jerzykiewicz et al.\ 2005, Handler
et al.\ 2006, Chapellier et al.\ 2006, De Cat et al.\ 2007 [hereinafter
DC07]). In addition, three subdwarf B stars were also discovered to show
"hybrid" oscillations (Oreiro et al.\ 2005, Schuh et al.\ 2006, Lutz et
al.\ 2009). The main physical difference between the B-type and A/F-type
"hybrid" pulsators is that in the first group the same driving mechanism
excites both types of oscillations, whereas in the
$\delta$~Sct/$\gamma$~Dor stars two different driving mechanisms are at
work.

A "hybrid" pulsator can therefore be characterized as a star that shows
(at least) two distinct sets of different self-driven pulsational mode
spectra, where the individual types of pulsations would belong to
established classes of variable stars. As known examples, this could be a
$\gamma$~Dor star that shows simultaneous $\delta$~Sct pulsations or a
$\beta$~Cep star that also shows SPB-type pulsations etc.

The different types of oscillations in a "hybrid" pulsator would be well
separated in frequency. Given the dispersion in the physical parameters
within the known classes of pulsators, a better discriminant would be the
pulsation constant $Q=P\sqrt{\bar\rho/\bar\rho_{\odot}}$, where $P$ is the
pulsation period and $\bar\rho$ is the mean stellar density. For instance,
the most evolved $\delta$~Sct stars show periods that fall into the range
of $\gamma$~Dor oscillations, but the Q values of the two groups are
cleanly separated (Handler \& Shobbrook 2002). Consequently, mixed-mode
pulsations by themselves are not "hybrid" pulsations because they occupy
the same frequency domain as pure p~modes.

The theoretical domains of B-type pulsators in the HR diagram are
sensitive to the input element mixture and opacity tables (e.g., Miglio,
Montalban \& Dupret 2007, Zdravkov \& Pamyatnykh 2008), a natural
consequence of the way their oscillations are driven. The driving is
caused by the huge number of the transitions inside the thin structure of
the electron shells in excited ions of the iron-group elements (Rogers \&
Iglesias 1994). Stankov \& Handler (2005) noticed that the observed
$\beta$~Cep instability strip appears tilted with respect to the
theoretically calculated boundaries, a problem nicely solved by models
with revised solar abundances (Asplund et al.\ 2004) or OP opacities
(e.g., Seaton 2005). As far as "hybrid" pulsators are concerned, the new
input physics predicts a much larger overlap region between $\beta$~Cep
and SPB stars in the HR diagram, and the frequency ranges of the excited
long and short period modes in "hybrid" B-type pulsators can pinpoint
which opacities are to be favoured (Miglio et al.\ 2007, Zdravkov \&
Pamyatnykh 2008).

Pulsational driving of $\delta$~Sct stars is due to the $\kappa$~mechanism
operating in the He{\sc ii} ionization zone (e.g., Chevalier 1971),
whereas the driving mechanism for $\gamma$~Dor stars is believed to be
convective blocking (Guzik et al.\ 2000, Dupret et al.\ 2004). The latter
mechanism is also expected to weaken or even exclude the driving of
$\delta$~Sct-type pulsations. However, as the instability domains of these
two types of variables overlap in the HR diagram (Handler 1999), "hybrid"
$\delta$~Sct/$\gamma$~Dor pulsators have been searched for, with success.

Despite all the possibilities that "hybrid" pulsators offer for
asteroseismology, the major observational problem that needs to be solved
before arriving at a unique seismic model still remains the same - or is
even more severe: a sufficiently large number of pulsation modes must be
detected and identified - in {\it both} frequency domains.

An even more basic observational problem that needs to be solved when
trying to discover "hybrid" pulsators is to prove that the variability,
again in {\it both} frequency domains, is intrinsic and due to pulsation.
For instance, some pulsating stars are the primary components of
ellipsoidal variables (e.g., Chapellier et al.\ 2004). Finally, care must
be taken that combination frequencies in frequency spectra are not
mistaken for intrinsic pulsation modes.

Consequently, the best targets for asteroseismic studies of "hybrid"  
pulsators need to be firmly established before embarking on large-scale
projects. To this end, we have selected six stars that have been claimed
as "hybrid" pulsators in the abovementioned literature for such an
exploratory study, comprising one main sequence A/F star and five O/B
stars. All these objects are bright enough and were expected to have
sufficiently large variability amplitudes to be observed with small
ground-based telescopes.

\subsection{Target stars}

The bright star $\gamma$~Peg (HD 886, B2\,IV) has been demonstrated to be
a $\beta$~Cep pulsator by Jerzykiewicz (1970). In the catalogue by Stankov
\& Handler (2005), it is the lowest-mass bona fide $\beta$~Cep star and it
is therefore close to the instability strip of the SPB stars (see Fig.\
1). Chapellier et al.\ (2006) reported two periods in the $\beta$~Cep
range plus two periods in the SPB star range from several spectroscopic
runs. Because neither of the published data sets on the star were
extensive, these authors suggested the acquisition of new and accurate
measurements. $\gamma$~Peg was suggested as the primary component of a
spectroscopic binary system. It is therefore possible that the slow
variations might be due to imperfect removal of the orbital radial
velocity variations. To confirm or reject the classification of
$\gamma$~Peg as a "hybrid" pulsator, a photometric study is ideal as no
correction for orbital motion needs to be made; the light time effect is
negligible.

HD~8801 (A7m) has been reported as a single Am star showing both
$\delta$~Sct and $\gamma$~Dor pulsations (Henry \& Fekel 2005). These
authors found two periods in the $\gamma$~Dor domain, two in the
$\delta$~Sct region, and two in between, with evidence for more.
Interestingly, the pulsation constants of the two intermediate periods of
HD~8801 are between the $\delta$~Sct and $\gamma$~Dor domains. This raises
the question whether or not HD~8801 has a companion, despite the lack of
orbital radial velocity variations reported by Henry \& Fekel (2005). If
it has not, then what is the physical cause of these intermediate-period
variations?

HD~13745 (V354~Per, O9.7\,II) and 53~Ari (HD~19374, B1.5\,V) have both
been reported as "hybrid" $\beta$~Cep/SPB stars by DC07 on the basis of
photometric data. When examining the frequency analyses for these two
stars (Figs.\ 21 and 22 of DC07), it becomes clear that these data suffer
from severe 1\,\cd aliasing and that the detection of the frequencies of
the light variations were made at low signal-to-noise. This combination
casts some doubt on the reality of the detections. We also note that while
53~Ari has a mass similar to $\gamma$~Peg and is thus close to the
$\beta$~Cep instability strip and to the SPB domain, HD~13745 has been
classified as an O-type star and may therefore the most massive "hybrid"
pulsator known to date.

$\iota$~Her (= HD~160762, B3\,IV) is the fifth "hybrid" pulsator candidate
in this study, again possibly sharing the pulsational properties of the
$\beta$~Cep and SPB stars. It has been most extensively studied by
Chapellier et al.\ (2000). These authors suggested that it is an SPB star
that is located close to the edge of the $\beta$~Cep instability strip
and that the claimed short-period variations may be of transient nature.
Therefore, some long-term monitoring was recommended. $\iota$~Her is also
a known wide spectroscopic binary, with a secondary star considerably less 
massive than the primary (Abt \& Levy 1978), and therefore unlikely to 
contribute to the system's recorded variability.

The claim of another "hybrid" pulsator, 53~Psc (= HD~3379, B2.5\,IV) has
been refuted because the suggested short-period variations could not be
confirmed (Le Contel et al.\ 2001, DC07). However, since 53~Psc is located
only a few degrees away from the primary target $\gamma$~Peg in the sky,
we decided to include it in this study.

\section{Observations}

We carried out differential time-series (u)vy photometry of the six stars
mentioned above with the 0.75-m Automatic Photoelectric Telescope (APT) T6
at Fairborn Observatory in Arizona, between October 2007 and June 2008. An
overview of the observations is given in Table~1.

\begin{table}
\begin{center}
\caption{Overview of the photometric observations. $\Delta T$ is the time
span of the data set in days, $T_{tot}$ is the total number of hours 
observed, $N_{tot}$ is the number of nights observed, and $N_{obs}$ is the 
number of data points obtained.}
\begin{tabular}{lccccc}
\hline
Star & $\Delta T$ & $T_{tot}$ & $N_{tot}$ & $N_{obs}$ & Filters \\
\hline
$\gamma$ Peg & 69 & 284 & 48 & 884 & uvy \\
53 Psc & 69 & 281 & 48 & 861 & uvy \\
HD 8801 & 69 & 272 & 48 & 737 & vy \\
HD 13745 & 69 & 243 & 47 & 616 & uvy \\
53 Ari & 69 & 283 & 48 & 927 & uvy \\
$\iota$ Her & 78 & 137 & 46 & 631 & uvy\\
\hline
\end{tabular}
\end{center}
\end{table}

As five targets are located at similar right ascension in the Northern
sky, they can be observed simultaneously. We used one local comparison
star for each target, four in total. The selection of comparison stars for
differential photometry is crucial to the success of the project and must
therefore be made carefully. In this work, the choice was based on two
primary reasons: the comparison stars should be unlikely to show
detectable variability, and they should, if possible, have been used in
previous work so that we may be able to verify their photometric constancy
within the accuracy of our measurements, and to compare our results with
those in the literature.

The comparison star for $\gamma$~Peg and 53~Psc was 34~Psc (HD~560,
spectral type B9Vn), already used by Sareyan et al.\ (1979), and suspected
to be a short-period variable by Jerzykiewicz \& Sterken (1990). Because
three stars were observed in this group, any possible variability in the
differential light curves can be unambiguously ascribed to the star
causing it. For HD~8801, we used HD~8671 (F7V) as local comparison. We
expected that this rather cool star would not show photometrically
detectable variability and it was used to examine our data for possible
effects of differential colour extinction. The local comparison star for
HD~13745 was HR~540 (HD~11408, A5m), as there was no suitable hotter star
in this part of the sky. Kurtz (1977) tested HR~540 for variability and
found it to be constant at the mmag level. The target 53~Ari was primarily
compared to $\sigma$~Ari (HD~17769, B7V), previously used as a comparison
star by Sterken (1988) and not found variable. The remaining target
$\iota$~Her was observed with the classical three-star method and with
respect to 77~Her (HD~158414, A4V) and 30~Dra (HD~162579, A2V), previously
observed by several authors (e.g., Chapellier et al.\ 2000, McMillan et
al.\ 1976) and never found variable.

We decided to observe in several filters, because this helps in
classifying the type of light variability (see Handler \& Shobbrook 2002
for a discussion). In addition, if pulsation is detected, mode
identifications can be obtained for variations with sufficient signal to
noise. We employed the Str\"omgren vy filters for all targets for high
precision and colour information plus the Str\"omgren u filter for
possible mode identification of the OB~type target stars. The integration
times were 40\,s in each filter for each star to keep photon and
scintillation noise below 1~mmag per measurement. A 10\,s sky integration
was obtained within each target/local comparison star group to take into
account the dependence of sky brightness on position of the Moon and air
mass.

Our observing strategy assured a minimum cadence of 25 minutes for
consecutive target star measurements; if some targets were out of reach of
the telescope, the cadence for the remainder was correspondingly higher.  
This enabled us to search for short periods because aliasing at the
sampling frequency was smeared out considerably.

The data were reduced following standard photoelectric photometry schemes.  
First, the measurements were corrected for coincidence losses. Then, sky
background was subtracted within each target/local comparison star group.
Extinction coefficients were determined on a nightly basis from the
measurements of all comparison stars via the Bouguer method (fitting a
straight line to a magnitude vs.\ air mass plot). The same extinction
correction was applied to each star. Finally, differential magnitudes were
computed by interpolation, the timings were converted to Heliocentric
Julian Date, and the data were subjected to analysis.

\section{Analysis and results}

The data were searched for periodicities using the program {\tt Period04}
(Lenz \& Breger 2005). This software package uses single frequency Fourier
and multifrequency nonlinear least squares fitting algorithms. The
analysis was started by computing the spectral window function and the
amplitude spectrum of the data, which were compared. If a signal was found
to be present at a significant level and supposed intrinsic, it was fitted
to the data, and its amplitude, frequency and phase were improved to
obtain an optimal fit. Consequently, this variation was subtracted from
the measurements (we call this procedure "prewhitening") and the residual
amplitude spectrum was computed and examined. If more than one periodic
signal was present, its parameters were optimized together with those of
the ones previously detected. Once no significant variation was left in
the residuals, the analysis was stopped. A signal was considered
significant when it exceeded a S/N ratio of 4, following the empirical
recommendation by Breger et al.\ (1993).

Before proceeding to the target stars, all possible differential light
curves of the comparison stars with respect to each other were examined.  
No evidence for intrinsic variability of any of the chosen comparison
stars was found within a level of $\sim 2$~mmag. Since the distances of
the different target/local comparison star groups on the sky was of the
order of tens of degrees, a 1/f component was present in most of the
amplitude spectra of the magnitude differences between the comparison
stars and some minor effects of differential (colour) extinction and/or
variable sky transparency cannot be ruled out. However, as can be seen
later, the differential light curves of the target stars had better
quality due to the use of local comparison stars.

Turning to the targets, the frequency search was first performed for the
light curves of each star in each filter individually. The solutions were
compared. Despite some variation in the S/N in the different filters, we
found excellent agreement between them. The light curves in the filter
with the highest S/N data were chosen as reference for the acceptance of
the reality of the signals. As the optimized frequency solutions varied
from filter to filter within the errors, we adopted S/N-weighted averages
as the final frequency values, fitted those to the light curves in the
individual filters and re-determined the amplitudes, phases and $S/N$ with
these fixed frequencies. Formal error estimates for these parameters were
determined with the formulae of Montgomery \& O'Donoghue (1999); the real
errors might be up to a factor of 2 higher (Handler et al.\ 2000,
Jerzykiewicz et al.\ 2005).

\subsection{$\gamma$ Peg}

The analysis of the amplitude spectrum of the y-filter data of 
$\gamma$~Peg is shown in Fig.\ 2. As variations occur in two frequency 
domains, where the spectral window function of the single-site data is 
different, we have computed it in a non-standard way: it is the amplitude 
spectrum of two signals of 0.63 and 6.59\,\cd with 5 mmag amplitude sampled 
in the same way as the original data (uppermost panel of Fig.\ 2). In this 
way, the aliases occurring at the reflection at frequency zero, which is 
important for the low-frequency domain, as well as the shape of the "pure" 
spectral window can both be displayed at once.

\begin{figure}
\includegraphics[width=88mm,viewport=05 00 280 545]{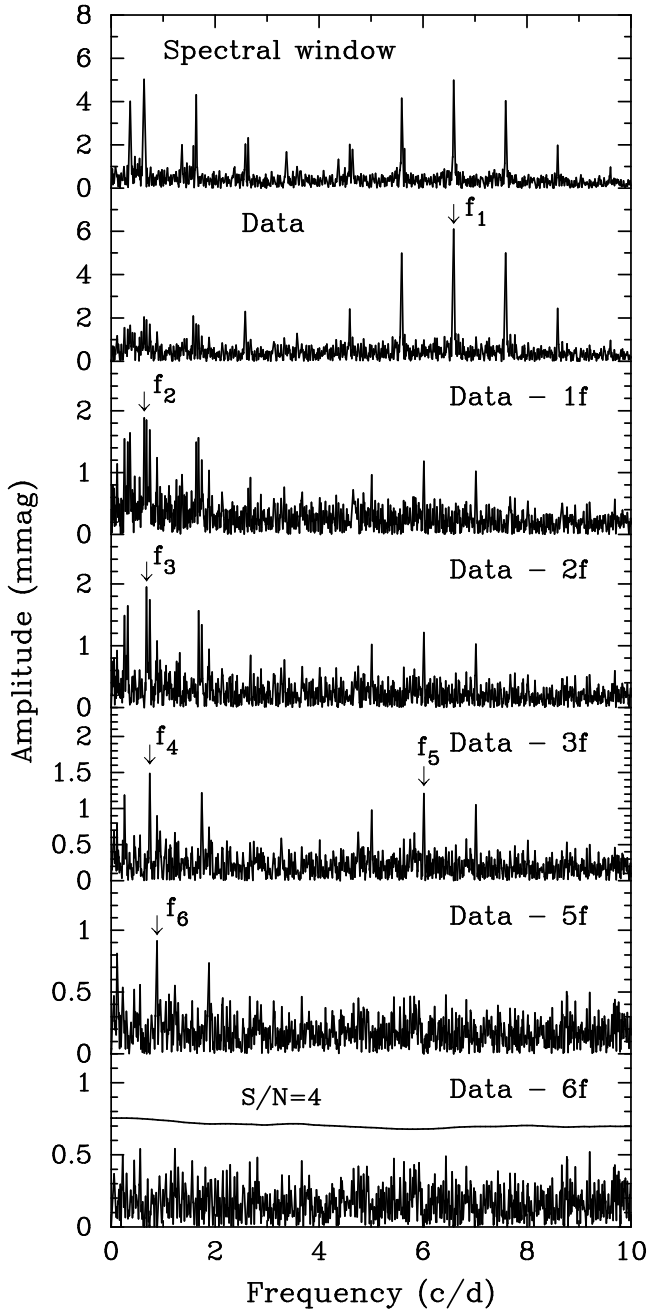}
\caption[]{Spectral window and amplitude spectra of the y-filter light 
curves of $\gamma$~Peg with consecutive stages of prewhitening. No 
variability was found outside the displayed frequency domain.}
\end{figure}

The amplitude spectrum of the data is dominated by the known 
$\beta$~Cep pulsation of $\gamma$~Peg. Prewhitening this signal leaves 
a small forest of low-frequency peaks that are all nicely resolved and 
unambiguously detected thanks to the high quality of our light curves (rms 
residuals of 4.0/3.2/2.7~mmag per point in $u/v/y$). 

We note that two frequencies separate in an amplitude spectrum if their
difference is larger than the inverse of the length of the data set and
that their amplitudes and phases can be determined free from systematic
errors if their separation exceeds 1.5 times the inverse of the length of
the data set (Loumos \& Deeming 1978). In the present case $1.5/\Delta T = 
0.022$\,\cd.

In the end, we could detect six independent signals in the light curves of
$\gamma$~Peg, four with frequencies around 0.7\,\cd and two with
frequencies around 6\,\cd (Table 2). The amplitudes of all these signals
increase with decreasing wavelength. No significant phase shifts between 
the different filters can be claimed for the individual oscillations.

\begin{table}
\begin{center}
\caption{Multifrequency solution for our light curves of $\gamma$~Peg. The
formal errors on the amplitudes are $\pm 0.1$ mmag in v and y and $\pm
0.2$ mmag in u. The S/N ratio and formal frequency uncertainty are
specified for the y filter data.}
\begin{tabular}{lccccc}
\hline
ID & Freq. & \multicolumn{3}{c}{Amplitude} & $S/N$ \\
 &  & u & v & y & \\
 & (\cd) & (mmag) & (mmag) & (mmag) & \\
\hline
$f_1$ & $6.5897 \pm 0.0002$ & 12.3 & 6.9 & 6.0 & 34.1 \\
$f_2$ & $0.6343 \pm 0.0005$ & 3.4 & 2.1 & 2.0 & 10.5 \\
$f_3$ & $0.6820 \pm 0.0005$ & 3.2 & 2.1 & 2.0 & 10.6 \\
$f_4$ & $0.7407 \pm 0.0007$ & 2.3 & 1.7 & 1.5 & 7.8 \\
$f_5$ & $6.0150 \pm 0.0009$ & 2.2 & 1.0 & 1.2 & 6.6 \\
$f_6$ & $0.8841 \pm 0.0011$ & 1.9 & 1.2 & 1.0 & 5.2 \\
\hline
\end{tabular}
\end{center}
\end{table}

\subsection{53 Psc}

There is clear evidence for variability in the amplitude spectrum of the
light curves of 53~Psc (second panel of Fig.\ 3). After prewhitening the 
strongest signal, a second variation with a frequency close to the first 
one becomes apparent. A two-frequency solution (Table 3) leaves no 
significant residual peak.

\begin{figure}
\includegraphics[width=88mm,viewport=05 00 280 335]{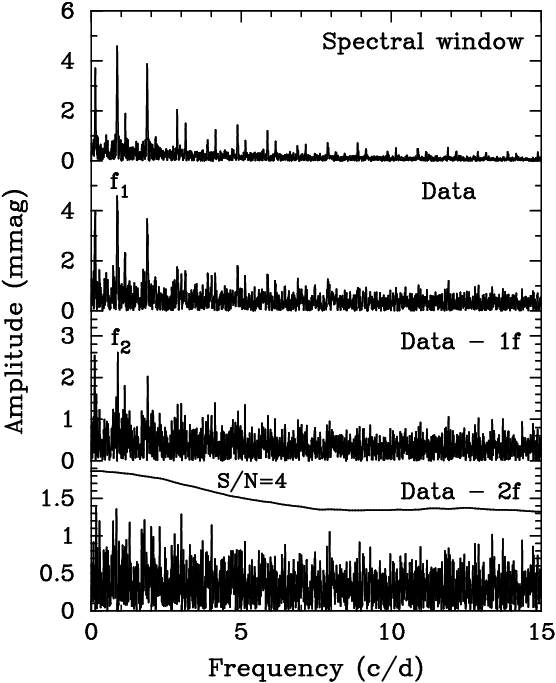}
\caption[]{Spectral window (computed as the amplitude spectrum of a 
single signal with a frequency of 0.863\,\cd and an amplitude of 4~mmag) 
and amplitude spectra of the u-filter light curves of 53 Psc with 
prewhitening of two closely-spaced frequencies. No variability
was found outside the displayed frequency domain.}
\end{figure}

\begin{table}
\begin{center}
\caption{Multifrequency solution for the light curves of 53 Psc. The 
formal errors on the amplitudes are $\pm 0.2$ mmag for all three filters. 
The S/N ratio and formal frequency uncertainty are specified for the u 
filter data. The signals are in phase within the observational errors.} 
\begin{tabular}{lccccc}
\hline
ID & Freq. & \multicolumn{3}{c}{Amplitude} & $S/N$ \\
 &  & u & v & y & \\
 & (\cd) & (mmag) & (mmag) & (mmag) & \\
\hline
$f_1$ & $0.8628 \pm 0.0005$ & 4.0 & 2.7 & 2.3 & 8.7 \\
$f_2$ & $0.8833 \pm 0.0007$ & 2.7 & 1.7 & 1.2 & 6.0 \\
\hline
\end{tabular}
\end{center}
\end{table}

However, the residual 53 Psc amplitude spectrum contains a peak at
0.84\,\cd at a S/N ratio of 3.7, which would form an equally spaced triplet
with the two detected frequencies. Additional measurements are required to
check its reality. The two detected signals have a
separation of about 1.4 times the inverse length of the data set, which
means they can be resolved, but our amplitude determinations may have some
small systematic errors. Nevertheless, the observed trend that the smaller 
the effective wavelength of the used filter the larger the amplitudes 
become is undoubtedly real.

\subsection{HD 8801}

The amplitude spectrum of the measurements of HD~8801 immediately confirms 
that three distinct time scales are present in the star's light 
variations. Therefore we have computed the spectral window of the data as 
the amplitude spectrum of an artificial light curve composed of three 
signals of 2.528, 7.976 and 21.669\,\cd with $y$ amplitudes of 3.2, 1.5 and 
1.3 mmag, respectively (upper panel of Fig.\ 4). Comparing this spectral 
window with the amplitude spectrum of the real data immediately shows that 
the variability in all three frequency regions is multiperiodic.

\begin{figure}
\includegraphics[width=88mm,viewport=05 00 280 545]{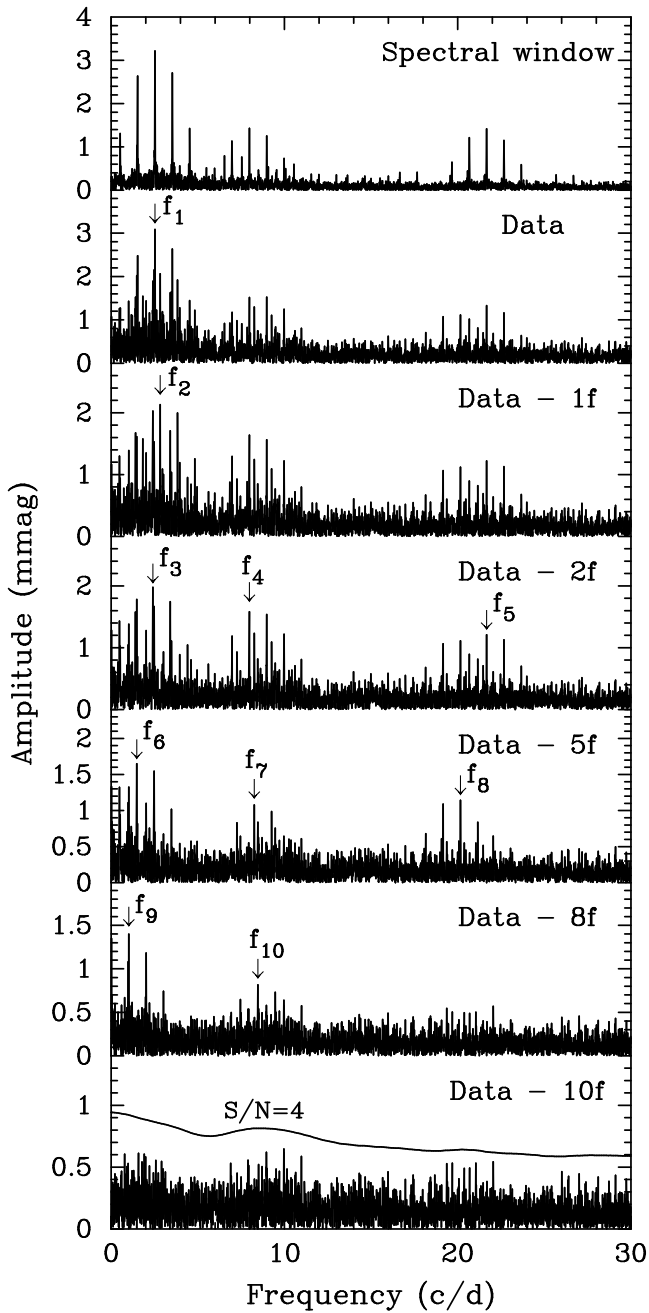}
\caption[]{Spectral window (see text for details on its calculation) and 
amplitude spectra of the y-filter light curves of HD 8801 with consecutive 
stages of prewhitening. No variability was found outside the displayed 
frequency domain.}
\end{figure}

\begin{table}
\begin{center}
\caption{Multifrequency solution for the light curves of HD~8801. The
formal errors on the amplitudes are $\pm 0.2$ mmag in v and $\pm 0.1$ mmag
in y. The S/N ratio and formal frequency uncertainty are specified for the
y filter data. The phase differences of the individual signals between 
the two filters are consistent with zero within 1.1$\sigma$.}
\begin{tabular}{lcccc}
\hline
ID & Freq. & \multicolumn{2}{c}{Amplitude} & $S/N$ \\
 &  & v & y & \\
 & (\cd) & (mmag) & (mmag) & \\
\hline
$f_1$ & $2.5277 \pm 0.0003$ & 4.4 & 3.2 & 14.6\\
$f_2$ & $2.8291 \pm 0.0005$ & 3.2 & 2.3 & 10.5\\
$f_3$ & $2.4057 \pm 0.0006$ & 3.0 & 2.0 & 9.1\\
$f_4$ & $7.9761 \pm 0.0007$ & 1.9 &  1.5 & 7.5\\
$f_5$ & $21.6692 \pm 0.0008$ & 2.0 & 1.3 & 8.4\\
$f_6$ & $1.4778 \pm 0.0005$ & 2.5 & 2.1 & 9.0\\
$f_7$ & $8.2607 \pm 0.0010$ & 1.5 & 1.1 & 5.3\\
$f_8$ & $20.1609 \pm 0.0009$ & 1.8 & 1.2 & 7.3\\
$f_9$ & $1.0178 \pm 0.0008$ & 2.1 & 1.4 & 6.3\\
$f_{10}$ & $8.4644 \pm 0.0013$ & 1.0 & 0.8 & 4.0\\
\hline
\end{tabular}
\end{center}
\end{table}

Successive prewhitening reveals the presence of ten independent and well
resolved frequencies in the light curves (Table 4); five between 1.0 and
2.8\,\cd, three between 7.9 and 8.5\,\cd and two between 20.1 and
21.7\,\cd.  The residual amplitude spectrum (lowest panel of Fig.\ 4)
suggests that additional signals are likely present in the light
variations. The $v/y$ amplitude ratios of all signals listed in Table~4
(between 1.19 and 1.52) are indicative of pulsation (cf.\ theoretical
predictions by, e.g., Garrido 2000).

None of the detected signals can be identified as having originated from
combination frequencies, even taking into account possible aliasing
ambiguities. The variation with frequency $f_9$ is difficult to be
interpreted due to its proximity to 1 cycle per sidereal day. It is also
possible that this signal is an alias from a long-term trend in the data,
resulting in a peak at 0.018\,\cd in the amplitude spectrum. However, the
frequency $f_9=1.0178$\,\cd given in Table 4 results in a slightly better
fit to the measurements.

\subsection{HD 13745}

This star shows some slow variability. Only a single frequency can be
detected at a significant level (Fig.\ 5 and Table 5). After subtracting
this signal, the highest peak in the residuals occurs at a frequency of
0.325\,\cd in all three filters. While this may be taken as evidence that
this signal is intrinsic, its does not reach a S/N ratio large enough for
a significant detection. Still, there is a strong 1/f component in the
residual amplitude spectrum that is not present in the residual
periodograms of the other targets that were measured in the same nights.  
This implies that more intrinsic variability should be present in our
measurements, but that we are unable to characterize it.

\begin{figure}
\includegraphics[width=88mm,viewport=05 00 280 260]{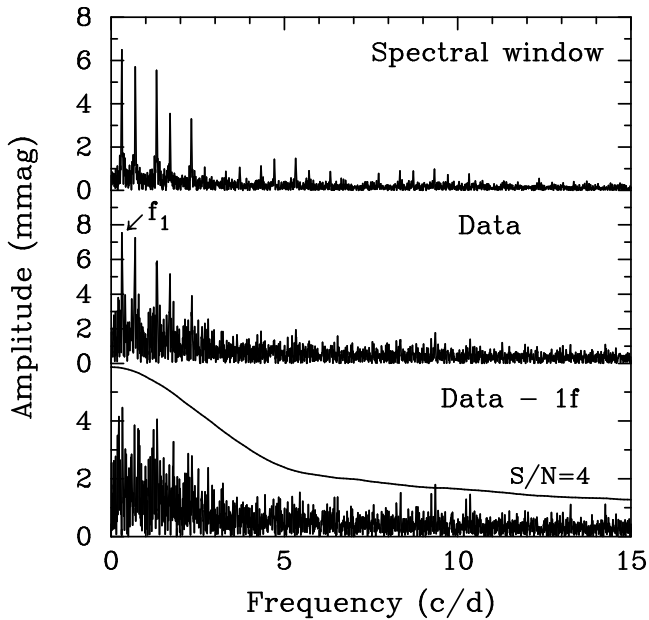}
\caption[]{Spectral window (computed as the amplitude spectrum of a single
signal with a frequency of 0.307\,\cd and an amplitude of 6~mmag) and
amplitude spectra of the y-filter light curves of HD 13745 with
prewhitening of one frequency. A formally significant peak near 9\,\cd
remains, but is not judged intrinsic (see text). No variability was found
outside the displayed frequency domain.}
\end{figure}

\begin{table}
\begin{center}
\caption{Frequency solution for the light curves of HD~13745. The 
formal errors on the amplitudes are $\pm 0.4$ mmag in the v and y filter, 
respectively, and  $\pm 0.5$ mmag in $u$. The S/N ratio and formal 
frequency uncertainty are specified for the y filter data. The 
variability is in phase in all three passbands within the observational 
errors.} 
\begin{tabular}{lccccc}
\hline
ID & Freq. & \multicolumn{3}{c}{Amplitude} & $S/N$ \\
 &  & u & v & y & \\
 & (\cd) & (mmag) & (mmag) & (mmag) & \\
\hline
$f_1$ & $0.3071 \pm 0.0005$ & 8.0 & 7.2 & 7.5 & 5.3 \\
\hline
\end{tabular}
\end{center}
\end{table}

Interestingly, some signals around 9\,\cd stand out of the noise in the
residual amplitude spectrum (lower panel of Fig.\ 5); one even reaches
formal significance. However, close inspection reveals that this peak
coincides with the 9\,\cd alias of the 0.325\,\cd signal mentioned above,
suggesting caution. Trial prewhitening of the signals near 9.3\,\cd does
not affect the ones near 0.3\,\cd, but prewhitening the 0.325\,\cd signal
does make the 9.3\,\cd peak disappear. We therefore conclude that these
higher-frequency signals are spectral window effects that reach formal
significance because of a lower influence of correlated noise in this
frequency domain. The question remains why this occurred in the 9\,\cd
region, and not in the frequency domain near 5\,\cd where aliasing is
stronger in this data set.

\subsection{53 Ari}

The analysis of our measurements of 53~Ari is difficult (Fig.~6).  There
are two formally significant signals in the amplitude spectrum, but both
are problematic. The strongest peak occurs close to 2 cycles per sidereal
day and may therefore an effect of atmospheric extinction. The second
signal that can be detected and is listed in Table~6 only reaches a
formally significant level after prewhitening the dubious $f_1$. Besides,
its amplitude is very low in the u filter and it is buried in the noise
there. We doubt that these two variations are intrinsic to 53~Ari and only
specify them in Table~6 for completeness and future reference.

\begin{figure}
\includegraphics[width=88mm,viewport=05 00 280 335]{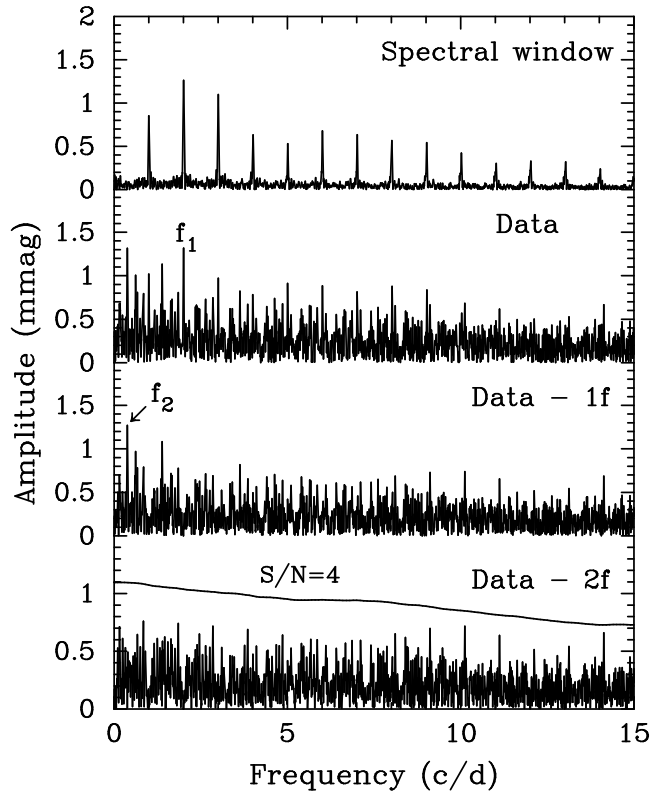}
\caption[]{Spectral window (computed as the amplitude spectrum of a single
signal with a frequency of 2.002\,\cd and an amplitude of 1.3~mmag)  
and amplitude spectra of the y-filter light curves of 53 Ari with 
prewhitening of two marginally significant frequencies. No variability
was found outside the displayed frequency domain.}
\end{figure}

\begin{table}
\begin{center}
\caption{Frequencies present in the light curves of 53 Ari. The
formal errors on the amplitudes are $\pm 0.1$ mmag in v and y and $\pm
0.2$ mmag in u. The S/N ratio and formal frequency uncertainty are
specified for the y filter data.}
\begin{tabular}{lccccc}
\hline
ID & Freq. & \multicolumn{3}{c}{Amplitude} & $S/N$ \\
 &  & u & v & y & \\
 & (\cd) & (mmag) & (mmag) & (mmag) & \\
\hline
$f_1$ & $2.0017 \pm 0.0009$ & 1.5 & 1.2 & 1.3 & 5.0 \\
$f_2$ & $0.3796 \pm 0.0009$ & 0.8 & 1.0 & 1.3 & 4.7 \\
\hline
\end{tabular}
\end{center}
\end{table}

\subsection{$\iota$ Her}

The amplitude spectrum of our measurements of $\iota$~Her contains no
significant peak (Fig.\ 7). The noise in this amplitude spectrum has the
same level and shape as that of the differential measurements of the
comparison stars. The tallest peak seen in this y-filter amplitude
spectrum is considerably less conspicuous than in $v$ and completely
absent in $u$. The behaviour of $\iota$~Her during our observations is
therefore consistent with non-variability.

\begin{figure}
\includegraphics[width=88mm,viewport=05 00 280 193]{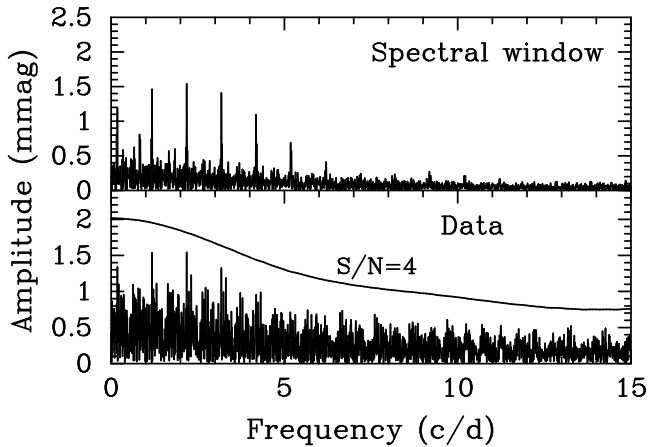}
\caption[]{Spectral window (computed as the amplitude spectrum of a single
signal with a frequency of 2.18\,\cd and an amplitude of 1.5~mmag)  
and amplitude spectra of the y-filter light curves of $\iota$~Her. No 
variability was found.}
\end{figure}

\section{Discussion and interpretation}

To understand the nature of the variability of the target
objects, knowledge about their positions in the HR diagram is required.  
These can be obtained by astrometric, photometric and spectroscopic means.
We have collected the corresponding literature data and results for all
our target stars.

Str\"omgren photometry indices are available for all targets from the
online version of The General Catalogue of Photometric Data (GCPD;
Mermilliod, Mermilliod \& Hauck 1997). We used the program by Napiwotzki,
Sch\"onberner \& Wenske (1993) that applies the $T_{\rm eff}$ calibration
by Moon \& Dworetsky (1985). Absolute magnitudes were determined from the
new reduction of the Hipparcos parallaxes (van Leeuwen 2007), whenever
significant. Otherwise, the $M_v$ calibration by Balona \& Shobbrook
(1984)  was used. With interstellar reddening determinations from
Str\"omgren photometry and bolometric corrections tabulated by Flower
(1996), we derived the stellar luminosities.

The GCPD also contains Geneva colour indices for all targets. We
determined their effective temperatures and surface gravities from these
data using the calibration by K\"unzli et al.\ (1997). DC07 used the same
procedure for the three stars we have in common, the difference being that
we used the values from the GCPD whereas they used the mean indices
derived from their own Geneva measurements. The differences in the results 
are negligible.

The parameters derived from these basic resources can be supplemented by
other work. Niemczura \& Daszy{\'n}ska-Daszkiewicz (2005) determined
log\,$T_{\rm eff} = 4.342\pm0.011$ and log\,$g=3.82$ for $\gamma$~Peg from
low-resolution ultraviolet spectra. A NLTE model atmosphere analysis from
optical spectra by Morel et al.\ (2006) resulted in log\,$T_{\rm eff} =
4.352\pm0.019$ and log\,$g=3.75\pm0.15$. These two spectroscopic results
are in excellent agreement, and they also agree with the outcome of the
calibrated photometry.

Concerning the Am star HD~8801, a spectral analysis of HD~8801 by R.\
Neuteufel (private communication) yields an effective temperature
of $\sim 7500$~K and a surface gravity consistent with that of a ZAMS
star. Spectroscopic results are also available for $\iota$~Her. Niemczura
(2003)  derived log\,$T_{\rm eff} = 4.251\pm0.005$ and log\,$g=3.84$ from
low-resolution ultraviolet spectra. Morossi et al.\ (2002) obtained
log\,$T_{\rm eff} = 4.251\pm0.017$ and log\,$g=3.64\pm0.25$ from visual
spectrophotometric measurements combined with ultraviolet spectra, and
Lyubimkov et al.\ (2002) determined log\,$T_{\rm eff} = 4.230\pm0.006$ and
log\,$g=3.77\pm0.12$ from a combination of several photometric and
spectroscopic temperature and gravity indicators, in good agreement with
previous literature values. For 53~Ari, Daszy{\'n}ska-Daszkiewicz (private
communication) determined log $T_{\rm eff} = 4.349$, log\,$g=3.55$ from
IUE spectra, without available error estimates.

The remainder of the target stars have no spectroscopic temperature or
surface gravity determinations. The only other constraint we found came
from Walborn (2002), who derived the absolute magnitude of HD 13745 from
its membership to the Per OB1 association with $M_v=-5.9$\,mag. No
additional data are available for 53~Psc.

The results of the different determinations of the basic parameters of our
targets quoted above are generally in good agreement, with one exception.
For the most luminous program star, HD~13745, DC07 listed an effective
temperature log\,$T_{\rm eff} = 4.550$ and log\,$g=4.03$ from Geneva
photometry but cautioned that this was obtained by extrapolation from
model atmosphere grids and may therefore be unreliable. However, these
values yields a luminosity consistent with the estimates derived by us,
although Str\"omgren photometry gives a considerably lower effective
temperature that we prefer to use in this work, also because it is more
consistent with the star's spectral luminosity classification. A
spectroscopic temperature determination for HD~13745 is desirable.

The stellar parameters adopted for the present work, as derived from 
combining the information quoted above, are summarized in
Table~7. The masses have been derived by placing the objects in a
theoretical HR diagram and by comparing their positions with stellar
models computed with the Warsaw-New Jersey stellar evolution code (e.g., 
see Pamyatnykh et al.\ (1998) for a description).

No current model input physics satisfy the conditions imposed by the
oscillation behaviour of all pulsating stars and their positions in the HR
diagram. Whereas OPAL opacities reproduce observed radial mode period
ratios for $\delta$~Sct stars better than OP opacities (e.g., Lenz,
Pamyatnykh \& Breger 2007), the latter provide a better match to the
excited frequency domains in $\beta$~Cep pulsators (e.g., Miglio et al.\
2007, Dziembowski \& Pamyatnykh 2008). Consequently, we used OPAL
opacities and the Grevesse \& Noels (1993) element mixture to model the
A/F star, but OP opacities and the Asplund et al.\ (2004) mixture for the
OB~stars. An overall metal abundance $Z=0.02$ and a hydrogen abundance of 
$X=0.7$ has been adopted for all models, and no convective core 
overshooting was used.

\begin{table}
\begin{center}
\caption{Basic parameters of the target stars derived from calibrated 
photometric systems, literature data and model evolutionary tracks. Error 
estimates have been conservatively adopted to include all possible 
sources of errors.}
\begin{tabular}{lccc}
\hline
Star & log~$T_{\rm eff}$ & log~$L$ & $M/M_{\odot}$\\
\hline
$\gamma$ Peg & $4.343\pm0.019$ & $3.77\pm0.20$ & $9.3\pm1.0$\\
53 Psc & $4.251\pm0.017$ & $3.11\pm0.15$ & $6.2\pm0.6$\\
HD 8801 & $3.866\pm0.009$ & $0.77\pm0.03$ & $1.54\pm0.03$\\
HD 13745 & $4.397\pm0.020$ & $5.10\pm0.25$ & $22.5\pm4.5$\\
53 Ari & $4.358\pm0.017$ & $3.47\pm0.20$ & $8.4\pm0.9$\\
$\iota$ Her & $4.246\pm0.017$ & $3.27\pm0.15$ & $6.7\pm0.6$\\
\hline
\end{tabular}
\end{center}
\end{table}

The locations of all target stars in the theoretical HR diagram are shown 
in Fig.\ 8, and are compared to observed (HD~8801) and theoretically 
predicted instability strips (OB star targets). All objects are inside 
the domains of "hybrid" pulsators, or no more than 2$\sigma$ away from 
these regions, and they will be discussed individually in what follows.

\begin{figure}
\includegraphics[width=88mm,viewport=00 05 285 515]{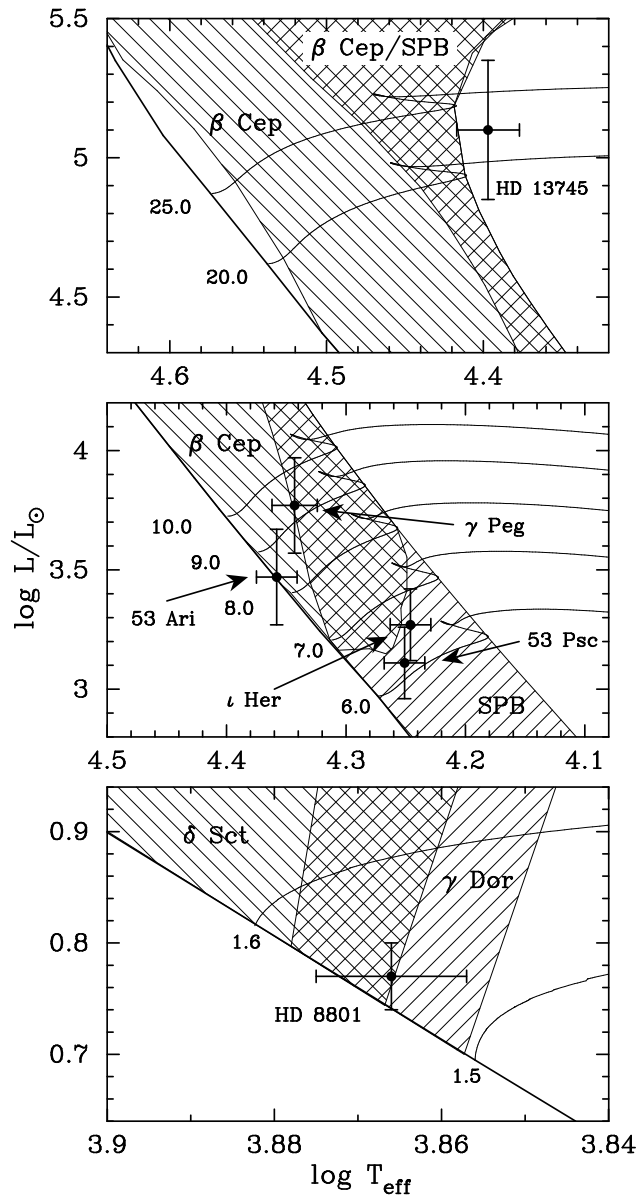}
\caption[]{Theoretical HR diagram in the physical parameter domains of our
target stars. The filled circles with the error bars denote the locations
of the targets. Some evolutionary tracks labeled with their masses on the
ZAMS are shown for comparison. Pulsational instability domains (Rodr\I guez
\& Breger 2001 for $\delta$~Sct stars, Handler \& Shobbrook 2002 for
$\gamma$~Dor stars, Zdravkov \& Pamyatnykh 2008 for $\beta$~Cep and SPB
stars) are indicated. Areas hatched from lower left to upper right are for
g-mode pulsators, areas hatched from lower right to upper left for p-mode
pulsators; regions of overlap consequently are cross-hatched.}
\end{figure}

\subsection{$\gamma$ Peg}

Our study confirms beyond any doubt that $\gamma$~Peg is a "hybrid"  
pulsator, as already suggested by Chapellier et al.\ (2006). In fact, the
agreement of these spectroscopic results with our photometric measurements
is surprisingly good, given the non-trivial analysis of the radial
velocities. Both $\beta$~Cep pulsation modes are present in the
spectroscopy as well as the highest-frequency SPB oscillation. The three
lowest-frequency SPB star pulsation modes are not resolved in the radial
velocities.

We have attempted mode identifications of the pulsational signals
detected. To this end, we followed the approach by Balona \& Evers (1999)
that uses theoretically calculated nonadiabatic parameters to determine
the amplitude ratios between different wavebands that are the
discriminator between the different modes of pulsation. We used the
evolutionary models computed in the previous section and calculated
theoretical uvy amplitudes for all pulsationally unstable modes along
model sequences that spanned the physical parameter space of $\gamma$~Peg
as listed in Table~7. Only those modes whose frequencies were in the
observed domain were considered. The resulting theoretical amplitudes in
the three passbands were normalized to that in the u filter, and the
average amplitude ratios and their rms errors were calculated. In this way
we can also estimate the uncertainties involved in the calculation of
theoretical results that need to be considered when being matched with the
observations.

Using the parameters for $\gamma$~Peg listed in Table~7, we arrived at the 
mode identification diagrams shown in Fig.~9. For modes $f_1$ and $f_5$, 
a frequency domain between 5.8 and 6.8\,\cd was scanned, for modes $f_2 - 
f_4$, we examined modes with frequencies of $0.62 - 0.76$\,\cd, and for 
mode $f_6$ the frequency range $0.86 - 0.91$\,\cd was searched. We chose 
these narrow frequency domains because the pulsation constants change 
considerably over the observed g~mode range, causing significant spread 
in the theoretical amplitude ratios (cf.\ Fig.\ 9).

\begin{figure*}
\includegraphics[width=170mm,viewport=-20 05 510 410]{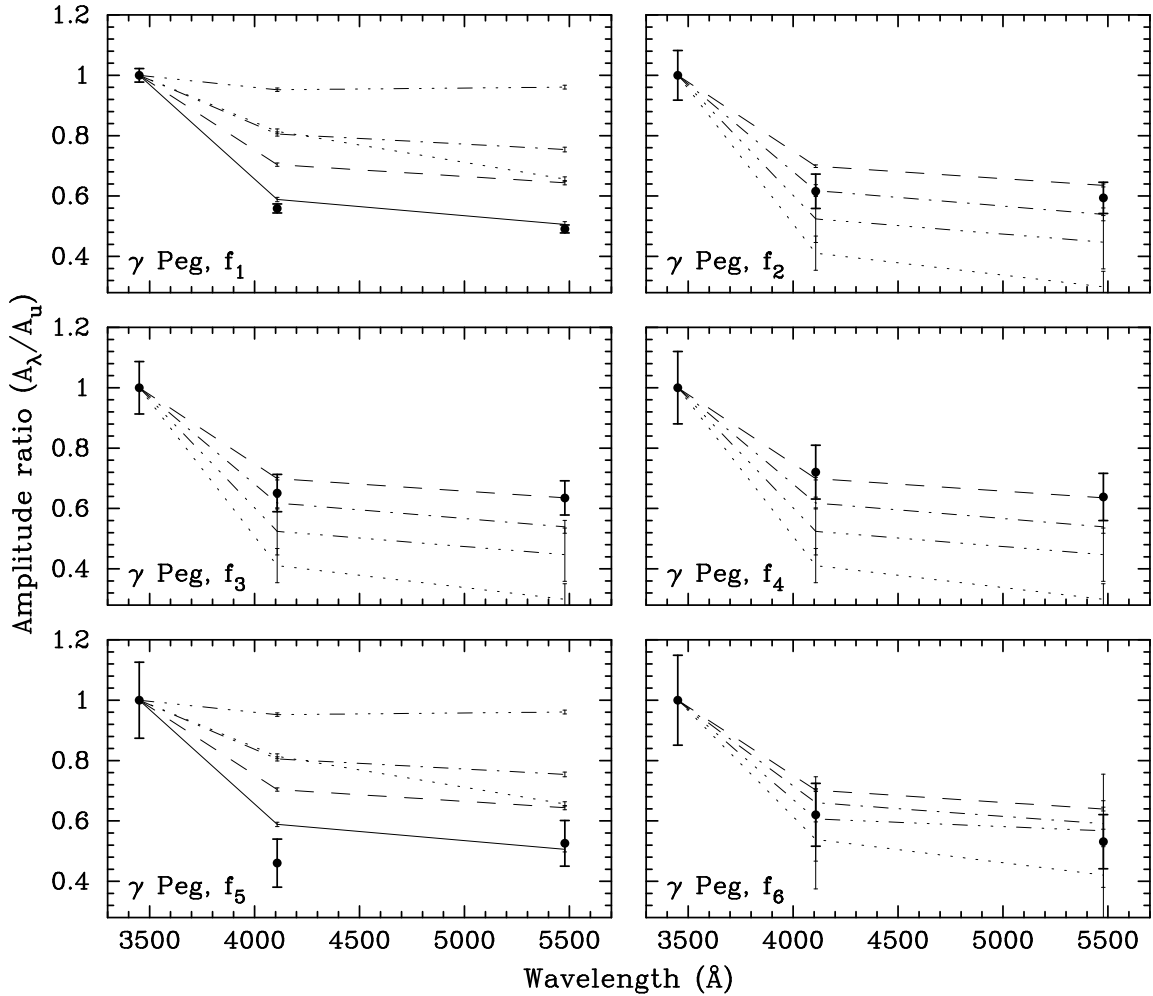}
\caption[]{A comparison of the observed amplitudes of $\gamma$~Peg in the
different filters with theoretical predictions of pulsational $uvy$
amplitude ratios, normalized to unity at $u$. The filled circles with
error bars are the observed amplitude ratios. The full lines are
theoretical predictions for radial modes, the dashed lines for dipole
modes, the dashed-dotted lines for quadrupole modes, the dotted lines for
$\ell=3$ modes, and the dashed-dot-dot-dotted lines are for $\ell=4$. The
thin error bars denote the uncertainties in the theoretical amplitude
ratios.}
\end{figure*}

We can interpret this diagram as follows: $f_1$ is certainly caused by a
radial mode. The mode $f_5$ is suggested to be radial from Fig.\ 9, but
the proximity of its frequency to $f_1$ ($f_5/f_1 = 0.913$) rules out such
an identification. $f_5$ must therefore be nonradial, and due to its
colour amplitudes would be most likely $\ell=1$. Concerning the SPB star
pulsation modes, $f_2$, $f_3$ and $f_4$ are most likely $\ell = 1$ or 2,
whereas no mode typing is possible for $f_6$ on the basis of the present
data.

Given the identification of $f_1$ as radial, consistent with previous
literature results (Stamford \& Watson 1977, Campos \& Smith 1980), and
given the near zero rotation of $\gamma$~Peg (e.g., see Stankov \& Handler
2005, Telting et al.\ 2006), we can try to identify $f_5$, and to restrict
the possible model parameter space for the star. We have a tight
constraint on the mean stellar density for any given radial overtone
corresponding to $f_1$, and we have a constraint on the stellar mass
(Table 7). Because $\beta$~Cep stars are main sequence objects, their
frequency spectra are sufficiently simple that one may hope that only a
few modes have a frequency corresponding to $f_5$ in the possible
parameter space.

We computed the corresponding models, again with the Warsaw-New Jersey
stellar evolution and pulsation code. Because we only have two pulsation
modes to fit, it is not possible to arrive at unique seismic models, but 
some constraints can be obtained.

For main sequence models with $9.3\pm1.0$\,M$_{\odot}$, three radial modes
can attain the observed frequency of $f_1=6.5897$\,\cd: the fundamental
mode as well as the first and the second overtone. Assuming the radial
mode to be the fundamental, we found that a model near 8.5\,M$_{\odot}$
has an $\ell=1$, g$_1$~mode at the observed frequency of $f_5=6.015$\,\cd.
If the radial mode was the first overtone, a $\sim 9.6$\,M$_{\odot}$
model has an $\ell=1$, p$_1$~mode at $f_5$. Would the radial mode be the
second overtone, an $\ell=2$, p$_2$~mode would match $f_5$ for a mass of
$\sim 9.4$\,M$_{\odot}$. However, we can rule out such an identification
from the colour amplitude ratios of $f_5$ in Fig.\ 9 that are inconsistent
with such an $\ell=2$ mode. Moreover, the corresponding model would have
an effective temperature 2.2$\sigma$ different from the value listed in
Table~7. No further model can reproduce the two $\beta$~Cep frequencies 
within the assumptions made.

We are therefore left with two possibilities for the position of
$\gamma$~Peg in the HR diagram. Interestingly, the mode frequencies
predicted by these two different models in the $\beta$~Cep domain are in
many cases similar. A distinction between these possibilities would
therefore require the detection (and possibly identification) of several
more of these pulsation modes that will, unfortunately, be of very low
amplitude. In both possible models, the observed g~modes would have radial
overtones between $\sim 8 - 12$ for $\ell=1$ and $\sim 14 - 21$ for
$\ell=2$, and thus are medium to high order g~modes typical of SPB stars.

\subsection{53 Psc}

The observational history in terms of variability of this star has been
nicely reviewed by DC07. In brief, $\beta$~Cep-type variations
were claimed by Sareyan et al.\ (1979), but later ascribed to originate
from its comparison star 34~Psc (Jerzykiewicz \& Sterken 1990). Le Contel
et al.\ (2001) suggested that multiperiodic SPB-type variations are
present in their spectroscopic measurements, but these could at best only
partly be confirmed by the photometric results of DC07.

We cannot confirm any of the literature results quoted above. Our data set
exceeds the previous studies both quantitatively and qualitatively; we
reached similar data quality as DC07, but have six times more
measurements. One of the frequencies claimed by Le Contel et al.\ (2001)
and DC07 is consistent with the 1\,\cd alias of the strongest signal we 
found, as a re-analysis of the measurements of DC07 suggests.

Our frequency $f_1$ for 53~Psc is present in the amplitude spectrum of
DC07's Geneva U~data. Prewhitening it from the Geneva data causes
disappearance of the strongest signal in this data set. We conclude that
it was due to an unfortunate coincidence between low signal to noise and a
poor spectral window. Because there is a time gap of several years between
the measurements by DC07 and ours, combining the two data sets with a
suitable choice of filters does not aid the analysis, that is, in any 
case, dominated by our measurements.

None of the other frequencies published by previous authors is present in
our data either. By comparison of the three possible differential light
curves between $\gamma$~Peg, 53~Psc and their comparison star 34~Psc we
can also exclude that 34~Psc showed short-period variations at the time
scales reported earlier, within a limit of 0.5\,mmag in the amplitude
spectrum of the combined uvy data. The frequencies $f_6$ found for
$\gamma$~Peg and $f_2$ found for 53~Psc are the same within the 
observational errors. However, the analysis of the differential light 
curves of the comparison stars exclude the possibility that a variation on 
this time scale is due to 34~Psc.

53~Psc appears to be a normal SPB star. Telting et al.\ (2006) found
high-degree line profile variations for this star, and DC07 interpreted
the variability they found in the same way. We attempted mode
identification of the two signals detected. Once more, we used the
Warsaw-New Jersey stellar evolution and pulsation code to compute models
in the parameter domain of interest (Table~7). We then calculated
theoretical amplitude ratios of g~modes with $1\leq \ell \leq 4$ in a
frequency range of $0.85 - 0.9$\,\cd and compare them with the
observations in Fig.\ 10.

\begin{figure}
\includegraphics[width=80mm,viewport=-10 00 266 281]{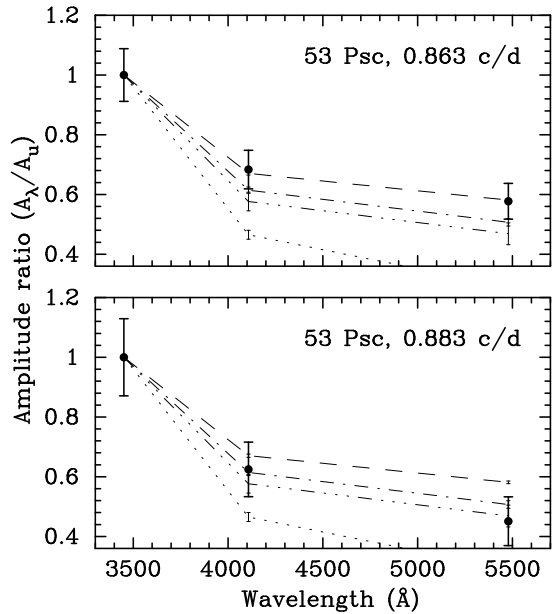}
\caption[]{A comparison of the observed amplitudes of 53~Psc in the
different filters with theoretical predictions of pulsational $uvy$
amplitude ratios, normalized to unity at $u$. The filled circles with
error bars are the observed amplitude ratios. The full lines are
theoretical predictions for radial modes, the dashed lines for dipole
modes, the dashed-dotted lines for quadrupole modes, the dotted lines for
$\ell=3$ modes, and the dashed-dot-dot-dotted lines are for $\ell=4$. The
thin error bars denote the uncertainties in the theoretical amplitude
ratios. The variability is consistent with pulsation in $\ell=1, 2$ or 4
modes.}
\end{figure}

The photometric amplitude ratios are consistent with pulsation with low
spherical degree $\ell$. We cannot rule out an $\ell=1$ identification as
DC07 did, although our error bars are smaller. This may be explicable with
the fact that photometric mode identifications for B stars critically
depend on ultraviolet amplitudes. Any systematic errors (as could for
instance be caused by low S/N data) on these amplitudes have an adverse
effect on the identifications. Finally, we remark that the high-degree
line profile variability reported by Telting et al.\ (2006) is unlikely to
correspond to the low-degree pulsations photometrically detected by us.

\subsection{HD 8801}

Given the limitations (single site observations) and differences (our time
base is a factor of $\sim 3.5$ larger) in the data sets, our results agree
well with those by Henry \& Fekel (2005). In the $\gamma$~Dor frequency
domain, their strongest signal is a combination of the 1\,\cd alias of our
signal $f_6$ and our frequencies $f_1$ and $f_3$ that were unresolved in
their data. The second strongest variation detected by Henry \& Fekel
(2005) in the $\gamma$~Dor frequency range is consistent with our results
within the errors. In the $\delta$~Sct domain, we find agreement with one
frequency and the 1\,\cd alias of the second, and in the intermediate
frequency domain we again find consistency keeping in mind the different
resolutions of the data sets.

We have attempted to identify at least the strongest modes with their
spherical degree $\ell$ and constructed amplitude ratio/phase shift
diagram as usually applied for $\delta$~Sct and $\gamma$~Dor stars
(e.g., see Watson 1988), for the Str\"omgren vy filters. We determined
theoretical areas of interest in such a diagram following Balona \& Evers
(1999), only to find out that the large error bars on the observational
amplitude ratios and phase shifts preclude any mode identification. Still,
the observed $v/y$ amplitude ratios strongly indicate that the variability
of HD~8801 in all three frequency domains is due to pulsation, as already
concluded by Henry \& Fekel (2005).

To compute the pulsation constants $Q$ of the individual oscillations, we
derived a mean density of $\rho= 0.46 \pm 0.12\, \rho_{\odot}$ from the
values in Table~7. Therefore, the $Q$ values have errors of $\sim$ 12\%.
For the modes with frequencies below 3\,\cd, $0.24<Q<0.67$, consistent
with high-order g-mode pulsation typical of $\gamma$~Dor stars. For the
two highest-frequency modes $0.031<Q<0.034$, consistent with low-order
p-mode and mixed-mode pulsation typical of $\delta$~Sct stars.

For the three signals with intermediate frequencies, however, we calculate
$0.080<Q<0.085$, corresponding to low-order g~modes ($k \sim 3$ for
$\ell=1$, $k \sim 5$ for $\ell=2$). These $Q$ values are {\it exactly} in
between the domains of $\delta$~Sct and $\gamma$~Dor stars (Handler \&
Shobbrook 2002), even taking into account the uncertainties determined
above. This variability is therefore either a previously unknown type of
stellar pulsation or it does not originate from HD~8801. The latter would
require a so far undetected optical companion that is a $\delta$~Sct star
near the TAMS, a rather unlikely scenario.

\subsection{HD 13745}

DC07 claimed the detection of "hybrid" $\beta$~Cep/SPB pulsation in this
star based on Geneva photometry. Our data set contains four times as many
measurements at about the same data quality. We find no evidence for
$\beta$~Cep pulsation of HD 13745, and we cannot confirm any of the
frequencies published by DC07. One of these frequencies
($f_3=0.6806$\,\cd) can be suspected to be an alias of the single
significant variation in our data, but a re-analysis of the Geneva U
measurements of DC07 does not support this idea. HD~13745 is therefore
variable on time scales around 3.2~d, with no evidence for "hybrid"
pulsations.

The parameters in Table~7 place the star somewhat beyond the TAMS in the
HR diagram (cf.\ Fig.\ 8). We computed theoretical pulsational amplitude
ratios for models between 17.5 to 27.5\,M$_{\odot}$ and compared them with
the observed uvy amplitudes (Table~5). The results are consistent with
pulsational variability. These models have modes excited at or near the
observed frequency of light variation. These are high-order g~modes (e.g.,
$k \sim 90$ for $\ell=1$ or $k \sim 150$ for $\ell=2$). However, since we
only detected a single frequency of light variation with certainty, this
still is no sufficient proof that the star is indeed a pulsator.

The published projected rotational velocity of HD~13745 is 260\,\kms (Hill
1967). Table~7 implies a stellar radius of $19\pm7$\,R$_{\odot}$. If the
0.3071\,\cd variation we found was due to rotation, a rotational velocity
of $295\pm110$\,\kms would result. This is consistent with the measured $v
\sin i$. We cannot exclude a rotational origin of the light variations at
this point; the uvy amplitudes and phases of the detected signal do
not provide additional clues. The excitation of r~modes is another
hypothesis that needs to be checked for HD~13745.

A conclusive interpretation of the variability of this star requires more
observational data, or measurements from space. In addition, a more
sophisticated approach to stellar modelling with respect to rotation
(e.g., Dziembowski, Daszy{\'n}ska-Daszkiewicz \& Pamyatnykh 2007), is
required as the time scales of the variability and rotation are of the
same order. Both these efforts are far beyond the scope of the present
work. The motivation for such studies may be drawn from the possible
relation of HD~13745 to supergiants that are suspected SPB stars on the
basis of investigations as suggested here (e.g., Saio et al.\ 2006).

\subsection{53 Ari}

The observational history of this object, classified as a runaway star by 
Blaauw (1956), in terms of variability, has been summarized by Sterken 
(1988). Photometric observations by this author did not reveal any light 
variations. Still, Gonzalez-Bedolla (1994) suggested the presence of
short-period variations with a time scale between 0.5 and 1\,hr.

"Hybrid" $\beta$~Cep/SPB pulsation of 53 Ari was reported by DC07, based
on 51 data points distributed over three years. Again, we cannot confirm
this result, and in this case, we have collected 18 times more
observations of comparable quality to DC07. Whereas we have formally
detected two low-amplitude signals in our light curves, we are not
convinced that these are intrinsic to the star (see Sect.\ 3.5). We
therefore do not examine this eventual variability further.

However, the discrepancies between the results of DC07 and ours, for all
three stars in common, require discussion. These authors reported the
detection of more oscillations than we did, on the basis of often
considerably smaller data sets. The difference lies in the use of
signal-to-noise criteria for frequency detection. We have followed the
suggestion by Breger et al.\ (1993) that signals with a S/N~ratio
exceeding 4 in an amplitude spectrum can be accepted as intrinsic, but
have been cautious when approaching this limit. DC07 used a S/N limit of
3.6.

The S/N~criterion by Breger et al.\ (1993) has been developed following an
examination of several data sets, checking the reproducibility of the
detection of the same frequencies in the same star. The test data were
however taken at multiple sites and had fairly clean spectral windows. In
most cases, the algorithms used for frequency analysis aim at a
minimization of the residual scatter in the measurements. In a data set
with limited temporal resolution and strong aliasing, such as single site
observations, the average level of the noise peaks interfering with the
spectral window function will therefore be reduced in the prewhitening
process, and the residual noise level will be underestimated.
Consequently, the significance of residual peaks in such an amplitude
spectrum will be overestimated. None of the variations claimed by DC07 for
the three stars we have in common exceeds an amplitude S/N ratio of 4 if
Breger et al.'s (1993) criterion is applied in the originally proposed
way. None of these variations is present in our measurements either,
despite our noise levels being between about 1/2 to 1/4 of the ones
reached by DC07.

We performed some tests as to whether or not the absence of the
frequencies published by DC07 for the three stars in common can be
explained by amplitude variations or beating or differences in detection
threshold. None of these frequencies could be found down to a S/N ratio of
only 2.5 (lower values were not considered) in our data.  Merging the
measurements by DC07 and/or the Hipparcos photometry of these stars with
ours did not confirm any of these frequencies because our data sets
dominate the combined analysis, no matter if using weights or not.

The width of the frequency interval in which the mean amplitude to
determine the noise level at the frequency of interest is computed is a
parameter to be carefully selected. It should be small enough to reflect
any change in the local noise level depending on frequency, yet large
enough to provide an accurate average: it must contain a sufficiently
large number of independent frequencies. The present work used a 5\,\cd
interval. Finally, it must be kept in mind that the amplitude spectrum
used to determine the noise level is oversampled. A frequency step of at
least $1/10\Delta T$, where $T$ is the length of the data set is required
to take the sinc shape of individual the peaks properly into account. This
work used a step of $1/20\Delta T$.

The discussion above illustrates why S/N~criteria or false alarm
probabilities must be used with particular caution when examining data
sets with complicated spectral window functions. We recall that Sect.\ 3.4
contains an example of a formally significant signal that was demonstrated
to be spurious.

Pulsational mode identification from signals detected at low $S/N$ in
photometry of B-type stars is also problematic. The theoretical pulsation
amplitudes of these stars increase towards the blue, as does the noise in
(ground based) observational data. Consequently, a spurious signal can be 
easily misidentified as being due to a pulsation mode when examining 
photometric amplitude ratios, as the amplitude will tend to be higher in 
the blue and ultraviolet.

\subsection{$\iota$ Her}

The most extensive study of the variability of $\iota$ Her has been
performed by Chapellier et al.\ (2000). These authors demonstrated the
presence of low-amplitude SPB-type variability with a time scale of a few
days both in photometry and in radial velocity, on top of orbital motion
with a 113\,d period. Short-period variability with a time
scale of $3 - 4$~hr was found in only one data set comprising seven 
consecutive nights of spectroscopy.

Our measurements could not confirm any of this variability. The absence of
slow variations can be explained with the amplitude changes of $\iota$ Her
reported by Chapellier et al.\ (2000). Turning to the short-term
variations, our detection level for periods between $3 - 4$~hr is around
1~mmag. The radial velocity amplitudes of the short-period signals found
by Chapellier et al.\ (2000) were below 0.3\,\kms, and quasi-simultaneous
photometric observations did not show variability on these time scales
either. Given the radial velocity to light amplitude ratios for main
sequence early B stars (e.g., Cugier, Dziembowski \& Pamyatnykh 1994), it
is well possible that these possible variations have escaped detection in
the aforementioned photometry and in our data set.

\section{Conclusions}

Asteroseismology of opacity-driven main sequence pulsators with their 
sparse mode spectra crucially depends on the correct determination of the 
intrinsic stellar pulsation frequencies and the correct identification of 
the oscillation modes causing them. Errors in this procedure will lead 
to incorrect seismic models, and incorrect conclusions about stellar 
physics will result.

This situation is to be avoided. Criteria that aid decisions which of the
observed frequencies are acceptable as intrinsic have their limitations,
and these have to be kept in mind. The final judgement of and
responsibility for data interpretation are the scientist's. In times with
ever improving quantities and quality of observational data it is
important that the results not be overinterpreted, even if potentially
exciting science provides temptation. Unfortunately, it usually takes more
effort and resources to dispute a result from the literature safely than
to make overoptimistic claims. Therefore, conservatism in data
interpretation is prudent.

In the literature on some of the B star targets of the present work,
transient short periods have been reported. In most cases, these came from
measurements poorly documented, and more extensive studies could never
confirm them. Meanwhile, a wealth of information is present on B-type
pulsators. Few of these stars show amplitude variability and if so, on
time scales of many years (Jerzykiewicz \& Pigulski 1999, Chapellier et
al.\ 2000). Beating of several pulsation modes can sometimes generate
single cycles of large-amplitude variability (which might just
occasionally exceed the observational detection threshold; see also the
discussion by Jerzykiewicz \& Sterken 1990), but if the target is
monitored intensively, at least some sign of them should be detectable in
a periodogram. A sudden, single occurrence of short-period variability in
a single, short data set of a well-monitored star otherwise quiescent on
these time scales is therefore doubtful. It is suggested that the
hypothesis of transient variability be set aside until convincingly 
proven.

The most interesting result from this study is the clear confirmation of
the suspected "hybrid" $\beta$~Cep/SPB pulsation of $\gamma$~Peg, with
four high-order g~modes and two low-order p or mixed modes detected. The
pulsation modes of the $\beta$~Cep type were used to restrict the
parameter space of possible models for the star in the HR diagram
considerably: the presence of a radial mode immediately constrained the
possible values for the mean density of the star. New measurements with a
higher sensitivity, preferably from space, may make detailed
asteroseismology of a "hybrid" pulsator possible for the first time.
Consequently, the star has been put forward as a target for the MOST
satellite (Walker et al.\ 2003).

We also confirmed that HD~8801 is a "hybrid" pulsator. In this object,
even three distinct groups of oscillation mode frequencies were detected.  
Two of those correspond to pulsations of known types, i.e.\ HD~8801 is
a $\delta$~Sct as well as a $\gamma$~Dor star. The third range of
excited modes is intermediate in frequency. If attributable to HD~8801,
these represent a previously unobserved type of stellar oscillations whose
excitation needs to be explained. An additional puzzle concerning "hybrid"  
$\delta$~Sct/$\gamma$~Dor stars is that all representatives known to date
are Am stars (Matthews 2007), although the incidence of pulsation in Am
stars is lower than in chemically normal A/F stars.

The only O star in our sample, HD~13745 showed slow, low-amplitude 
variability. Since this star may have left the main sequence, it could be 
related to the recently postulated supergiant SPB stars (Saio et al.\ 
2006). This hypothesis requires confirmation. The most efficient strategy 
would be a high-resolution spectroscopic study, serving two purposes: to 
place the star in the HR diagram reliably, and to test whether or not the 
variability is indeed due to pulsation.

This study was unable to shed new light on the behaviour of $\iota$~Her.  
We failed to detect photometric variations during the observations, which
might have been below our detection threshold. To understand the
variability of this star, only a study involving simultaneous space
photometry and high-resolution ground-based spectroscopy may suffice.

Concerning the two remaining target stars, 53 Psc was shown to be an SPB 
star. Two pulsation modes were detected, and some additional oscillations 
may be hidden in the observational noise. There is no evidence that this 
star also shows short-period oscillations reminiscent of a $\beta$~Cep 
star. The same statement is true for 53~Ari that did not 
show convincing variability on longer time scales either.

\section*{ACKNOWLEDGEMENTS}

This work has been supported by the Austrian Fonds zur F\"orderung der
wissenschaftlichen Forschung under grant P20526-N16. The author is
grateful to Peter De Cat for providing his observational data and for
discussing the frequency analyses of DC07. Luis Balona, Wojtek Dziembowski
and Alosza Pamyatnykh are thanked for permission to use their codes.
Patrick Lenz provided some pulsation models during the earlier stages of
this work. Vichi Antoci, Conny Aerts, Jadwiga Daszy{\'n}ska-Daszkiewicz
and Alosha Pamyatnykh supplied several helpful comments on the manuscript,
and the constructive remarks of the anonymous referee made this paper more
complete.

\bsp

\end{document}